\documentclass[12pt]{article}
\usepackage{amsmath,amssymb,graphicx}
\makeatletter \@addtoreset{equation}{section}

\usepackage{cite}
\usepackage{bm}
\usepackage{dcolumn}
\makeatother
\def\one{{\hbox{ 1\kern-.8mm l}}}
\newcommand{\Dslash}{\not{\hbox{\kern-4pt $D$}}}
\newcommand{\pdslash}{\not{\hbox{\kern-2pt $\partial$}}}
\newcommand{\be}{\begin{equation}}
\newcommand{\bea}{\begin{eqnarray}}
\newcommand{\eea}{\end{eqnarray}}
\newcommand{\ba}{\begin{array}}
\newcommand{\ea}{\end{array}}
\newcommand{\ee}{\end{equation}}

\begin{document}

\begin{titlepage}
\vspace*{1mm}%
\hfill%
\vbox{
    \halign{#\hfil        \cr
           IPM/P-2010/050 \cr
          % SUT-P-07-2b   \cr
                     } % end of \halign
      }  % end of \vbox
\vspace*{15mm}%
\begin{center}

{{\Large {\bf P-Wave Holographic Insulator/Superconductor Phase Transition }}}

\vspace*{15mm} \vspace*{1mm} {Amin Akhavan$^{a,b}$ and Mohsen Alishahiha$^{a}$}

 \vspace*{1cm}

{\it ${}^a$ School of physics, Institute for Research in Fundamental Sciences (IPM)\\
P.O. Box 19395-5531, Tehran, Iran \\ }

\vspace*{.4cm}

{\it ${}^b$ Department of Physics, Sharif University of Technology \\
P.O. Box 11365-9161, Tehran, Iran}

\vspace*{.4cm}

email: amin\_akhavan@mehr.sharif.ir, and  alishah@ipm.ir

\vspace*{2cm}
%%\maketitle
\end{center}

\begin{abstract}
Using a five dimensional AdS soliton in an  Einstein-Yang-Mills theory with $SU(2)$ gauge 
group we study p-wave holographic insulator/superconductor phase transition. To explore
the phase structure of the model we consider the system in the probe limit as well
as fully back reacted solutions. We will also study  zero temperature limit of the 
p-wave holographic superconductor in four dimensions.

\end{abstract}

\vspace*{3cm}

\begin{center}
{\it Dedicated to the memory of   Mohammad Hossein Sarmadi}
\end{center}

\end{titlepage}

\section{Introduction}

The application of  AdS/CFT correspondence \cite{Maldacena:1997re} in condensed matter physics  has been attracted lots of attentions in recent years.
In particular it has provided gravitational descriptions for systems which exhibit  superconductor/superfluid\cite{{Gubser:2008px},{Hartnoll:2008vx},{Hartnoll:2008kx}} phases.
Since in  condensed matter physics  we are typically dealing with  systems at
finite charge and temperature, in the context of the AdS/CFT correspondence the dual gravity
 descriptions should be given by gravitational models which admit charged black holes as 
vacuum solutions.  

Indeed the simplest model may be provided by an  Einstein-Maxwell theory coupled to
a charged scalar field. For this model, it has been shown that the charged black holes become
unstable to develop scalar hair for sufficiently low temperature \cite{Gubser:2008px}
and moreover  the $U(1)$ symmetry is broken near the black holes horizon.
Within the framework of the AdS/CFT correspondence, the charged scalar field 
corresponds to an 
operator which carries the charge of the global $U(1)$ symmetry. Having non-zero hair 
corresponds to the fact that the
dual operator has non-zero expectation value and the global $U(1)$ symmetry is
 broken as well.  This phenomena may be interpreted as a second order phase transition
between   conductor and superconductor phases. This interpretation  has been  
supported by making use of the behavior of  AC conductivity in these phases.

Using a five dimensional AdS soliton in an Einstein-Maxwell-charged scalar field
the authors of \cite{Nishioka:2009zj} have constructed a model describing 
an insulator/superconductor phase transition at zero temperature. Actually since in this model 
the normal phase is described by an AdS soliton where the system exhibits  
mass gap\cite{Witten:1998zw} the dual field theory is in  an insulator phase.
On the other hand for sufficiently large chemical potential the AdS soliton becomes
unstable to forming scalar hair which corresponds to the fact that  the dual theory 
is in a superconductor phase. Holographic insulator/superconductor system with back reactions has also been 
studied in \cite{Horowitz:2010jq}.

The aim of the present article is to extend the holographic insulator/superconductor for the case
where the AdS soliton develops a vector hair. In fact  holographic superconductors with
vector hair, known as p-wave holographic superconductors, have been first studied in  
\cite{Gubser:2008zu} and further explored in \cite{{Gubser:2008wv},{Roberts:2008ns}}.
The simplest example of  p-wave holographic superconductors may be provided by
 an Einstein-Yang-Mills theory with $SU(2)$ 
gauge group\footnote{See also \cite{Aprile:2010ge} for p-wave holographic
 superconductors in the context of  five dimensional gauged supergravity.}. 
In this model the electromagnetic $U(1)$ gauge symmetry is identified with the 
abelian  $U(1)$ subgroup of the $SU(2)$. The other components of the $SU(2)$ gauge
field play the role of charged fields whose non-zero expectation values break
 the $U(1)$ symmetry leading to a phase transition in the dual field 
theory.

In this paper following \cite{Gubser:2008zu} we  will  consider a five 
dimensional  $SU(2)$ Einstein-Yang-Mills theory with a negative 
cosmological constant whose action
 is given by\footnote{In our notation we have set the five dimensional
Newton constant to one, $\kappa_5=1$.}
\be\label{action}
S=\int d^5x\sqrt{-g}\left[\frac{1}{2}\left(R-\Lambda\right)-\frac{1}{4}F^a_{\mu\nu}F^{a\;\mu\nu}\right],
\ee
where $F_{\mu\nu}$ is the field strength of the $SU(2)$ gauge field. In our notation the  negative cosmological constant is given by $-12/L^2$ with $L=1$.   The equations of motion
coming from the above action are
\bea\label{eom}
R_{\mu\nu}-\frac{1}{2}g_{\mu\nu} R-6g_{\mu\nu}= T_{\mu\nu},\;\;\;\;\;
\frac{1}{\sqrt{-g}}\partial_\mu(\sqrt{-g}F^{a\;\mu\nu})+qf^{abc}A_\mu^b F^{c\;\mu\nu}=0,
\eea
where 
\be
T_{\mu\nu}=F_{\mu\rho}^aF^{a\;\rho}_{\nu}-\frac{1}{4}g_{\mu\nu}
F_{\mu\nu}^aF^{a\;\mu\nu},
\ee
with $F_{\mu\nu}^a=\partial_\mu A^a_\nu-\partial_\nu A^a_\mu+qf^{abc}A^b_\mu A^c_\nu$.

The model has only  one free parameter which in our notation is given by the gauge
coupling $q$. Whether or not the back reactions of the gauge field on the metric are
important is  controlled by $q$. When $q$ is large the effects of the gauge field on 
the geometry are negligible, while for finite $q$ the gauge field back reactions are 
important and affect the geometry. 

The equations of motion \eqref{eom} support an AdS solitonic solution with zero  gauge field 
and the metric which is given  by 
\bea\label{metric}
ds^2=\frac{1}{r^2g(r)}dr^2+r^2(-dt^2+dx^2+dy^2)+r^2g(r) d\chi^2,
\;\;\;\;\;\; 
g=1-\frac{r_0^4}{r^4}.
\eea
Essentially this solution can  be obtained from a five dimensional AdS black hole solution
by making use of two Wick rotations. Physically, this solution corresponds to a five 
dimensional solution compactified on a circle with anti-periodically boundary condition
for the fermions along the compact direction. Since the geometry ends at $r=r_0$ where $g_{tt}$ is non-zero, the background provides the  gravity description of a three dimensional field theory with a mass gap\footnote{We would like to thank referee for his/her
comment on this point}.

It is worth mentioning that unlike the AdS black hole that finiteness of the gauge potential
at the horizon prevents to have a non-zero $A_t$ at the horizon, in the present case one could still have the above solution with a constant non-zero gauge potential $A_t=\mu$. 

The theory given by the action \eqref{action} admits another analytic  solution with non-zero gauge field. The corresponding 
solution is, indeed, an  AdS Reissner-Nordstr\"om black hole which carries 
the charge of the $U(1)$
abelian subgroup of the $SU(2)$ gauge group. The solution  is given by
\be\label{RN}
ds^2=\frac{dr^2}{g}-gdt^2+r^2(dx^2+dy^2+dz^2),\;\;\;\;\;\;\;\;\;
A=\rho\left(1-\frac{1}{r^2}\right)\sigma^3 dt,
\ee
where $g=r^2-\frac{1+\rho^3/3}{r^2}+\frac{\rho^2}{3r^2}$, and
$\sigma^3$ is the generator of $U(1)$ subgroup\footnote{We denote the  generators of 
$SU(2)$ by $\sigma^i$ for $i=1,2,3$.}. Here we have normalized the coordinates such 
that the horizon is located at $r=1$. In this notation the  Hawking temperature of the
black hole is $T=\frac{2-\rho^2/3}{2\pi}$.

The organization of this paper is as follows. In the next section we study p-wave
holographic insulator/superconductor in the probe limit where we will also
compute the AC conductivity of the model. In section three we first consider fully  
back reacted solutions and then we study the phase structure of the theory. 
In section four, for completeness, we will study zero temperature limit of p-wave 
holographic superconductor in four dimensions.The last section is 
devoted  to conclusions. 

%%%%%%%%%%%%%%%%%%%%%%%%%%%%%%%%%%%%%%%%%%%%%%
%%%%%%%%%%%%%%%%%%%%%%%%%%%%%%%%%%%%%%%%%%%%%%

\section{Insulator/Superconductor transition at probe limit}

In this section we will consider dynamics of the $SU(2)$ gauge
field on the background\eqref{metric}. 
In general turning on a gauge field, the background metric \eqref{metric}
gets corrections due to back reactions of  the gauge field back. 
Nevertheless at first order we will 
consider the case where the back reactions are  negligible so that the gauge field may be
treated as a probe. This can be done with the assumption that $q$ is sufficiently 
large \cite{{Gubser:2008wv},{Roberts:2008ns}}.

%%%%%%%%%%%%%%%%%%%%%%%%%%%%%%%%%%%%%%%%%%%%%%%

\subsection{Phase transition}

To study the insulator/superconductor phase transition, following\cite{Gubser:2008wv}, we 
will consider the  following  ansatz for the gauge field\footnote{See also
\cite{{Basu:2008bh},{Ammon:2008fc}}.}
\be
A=\phi(r)\;\sigma^3 dt+\psi(r)\;\sigma^1 dx.
\ee
Since we are in the probe limit, the back reactions of this
gauge field on the metric \eqref{metric} are negligible. Note that  in this ansatz the 
$t$-component of the gauge field represents the $U(1)$ gauge field, while the second term
plays the role of the charged field whose condensation breaks the   $U(1)$ gauge symmetry.

Plugging this ansatz into the equations of motion \eqref{eom} one arrives at 
\bea\label{equations}
\phi''+\left(\frac{3}{r}+\frac{g'}{g}\right)\phi'-\frac{\psi^2}{r^4 g}\phi=0,\;\;\;\;\;
\psi''+\left(\frac{3}{r}+\frac{g'}{g}\right)\psi'+\frac{\phi^2}{r^4 g}\psi=0.
\eea
The aim is to solve the above equations, though in general it is difficult to do it 
analytically and therefore we will utilize the numerical method. To proceed we note that
the above equations are invariant under the rescaling $r\rightarrow r_0 r,\phi\rightarrow r_0\phi$ and $\psi\rightarrow r_0\psi$, so that  $r_0$ may be dropped from
 the equations and leaves  us to work with dimensionless quantities. 
 
Near the boundary where $r\rightarrow \infty$ the solutions of the equations behave as
\be
\phi=\mu-\frac{\rho}{r^2},\;\;\;\;\;\;\;\;\;\;\psi=\psi_0+\frac{\psi_1}{r^2}.
\ee
From dictionary of the AdS/CFT correspondence\cite{{Gubser:1998bc},{Witten:1998qj}},
It is known that $\mu$ and $\rho$
correspond to the chemical potential and charge density in the boundary theory, while $\psi_0$ and $\psi_1$  may be identified  as a source and the expectation value of the dual operator,
respectively. Since we are interested in the case where the dual operator is not sourced, we set
$\psi_0=0$.  Also up to a normalization one has $\langle O\rangle\sim \psi_1$.

\begin{figure}
\begin{center}
%\vspace{1cm}
\includegraphics[height=4.86cm, width=8cm]{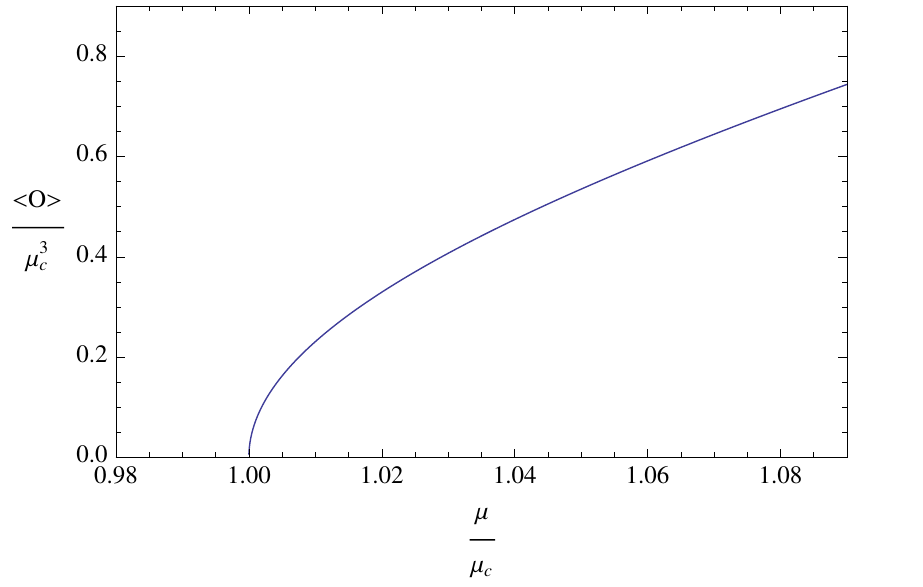}
\includegraphics[height=5cm, width=7.6cm]{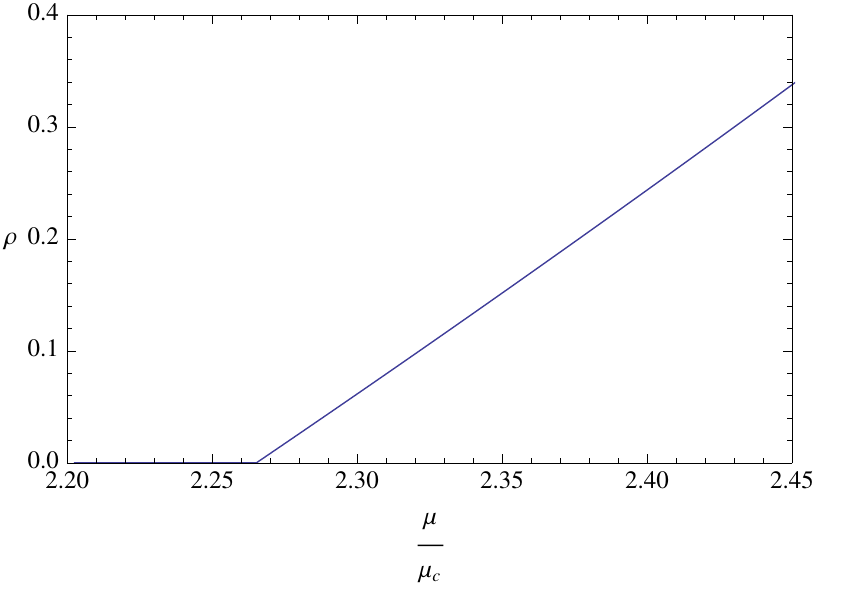}
\caption{ Here we plot the behaviors of $\langle \cal{O}\rangle$  and charge density $\rho$
for various values of chemical potential $\mu$. Here $\mu_c=2.26$.} \label{fig1}
\end{center}
%\vspace{0.5cm}
\end{figure}

On the other hand imposing the finiteness condition near the tip where $r\rightarrow 1$, 
leads to the following series expansions for the $\psi$ and $\phi$ fields (see also  \cite{Nishioka:2009zj})
\bea
\psi&=&\alpha_0+\alpha_1 (1-\frac{1}{r})+\alpha_2 (1-\frac{1}{r})^2\cdots,\cr &&\cr
\phi&=&\beta_0+\beta_1(1-\frac{1}{r})+\beta_2(1-\frac{1}{r})^2\cdots\,.
\eea
With these boundary conditions one may solve the equations \eqref{equations} to find 
the expectation value of the dual operator as a function of the chemical potential.
In fact given the initial values for the gauge  field near the tip, one may find the boundary values
of the gauge field by making use of the shooting method.  
Indeed doing so, one finds that the solution is unstable 
to develop a hair for the chemical potential bigger than a critical value, $\mu >\mu_c$.
More precisely using the Mathematica we have numerically solved the equations 
leading to the plot 
depicted in figure \ref{fig1}. In our numerical method we have chosen $\phi$ at the tip as the shooting parameter and fixed $\psi$ at the tip by making use of the asymptotic condition $\psi_0=0$. 

From the figure \ref{fig1} (right) we observe that the expectation value of the dual operator is
non-zero for $\mu>\mu_c=2.26$. Having non-zero expectation value spontaneously breaks the 
$U(1)$ symmetry  leading to a phase transition. To find the nature of the 
phase transition, it is illustrative  to find the behavior of the charge density in terms of the chemical potential. Indeed since  the charge density may be identified with the derivative of free energy, $\rho=\partial F/\partial\mu$,  its behavior around the critical chemical potential can
indicate the nature of the phase transition. Indeed, as it is evident from figure \ref{fig1} (right), it turns out that the phase transition is second order. In this sense the behavior of the
 system is  similar to that for  the s-wave one\cite{Nishioka:2009zj}.

Since the geometry \eqref{metric} is dual to a three dimensional theory with a mass gap, it is 
natural to treat  the phase of the system before condensation as an insulator phase. Therefore the obtained phase transition may be interpreted as an insulator/superconductor
phase transition. To see whether the new phase behaves as a superconductor  it is useful 
to study the response of the theory to an external magnetic field. In other words one may
compute the AC conductivity of the system.  

\subsection{AC conductivity}

To study the conductivity of the theory  we will consider an extra magnetic field along  
$y$ direction. From gravity description point of view this can be done by turning on 
a non-zero gauge field in  $y$ direction. To do so, one may consider the following 
ansatz for the gauge field along $y$ direction
\be
A_y=A(r)e^{i\omega t}\sigma^3.
\ee
Since  we are working in the probe limit,  we will consider the back reactions
of the new component of the gauge field neither on the metric nor on the other components
of the gauge field. As the result it is sufficient to consider the equation of motion for
 $A_y$  in the background generated by the metric \eqref{metric} together with the solutions
of $\psi$ and $\phi$ given by the equations \eqref{equations}. 

The corresponding equation of motion for $A_y$ is given by
\be\label{eqA}
A''+\left(\frac{3}{r}+\frac{g'}{g}\right)A'+\frac{\phi^2}{r^4 g}\left(\omega^2-\psi^2\right)A=0.
\ee
For large $r$ where we approach the boundary the behavior of the gauge field is 
\be
A=A_{0}+\frac{A_{1}}{r^2}.
\ee
With this notation the conductivity in $y$ direction is given by (see for example \cite{Roberts:2008ns})
\be
\sigma_{yy}=\frac{-i A_{1}}{\omega A_{0}}.
\ee
Using the numerical solution we had found for $\psi$ and $\phi$ in the previous
 subsection we can find the behavior of the AC conductivity in terms of the energy $\omega$.
Actually we expect that the imaginary part of the conductivity to have 
a non trivial behavior at $\omega=0$. More precisely we would expect to get a delta function
support for the real part of the conductivity at $\omega=0$. Indeed this is what we get from our numerical solution. 

\begin{figure}
\begin{center}
%\vspace{1cm}
\includegraphics[height=6cm, width=9cm]{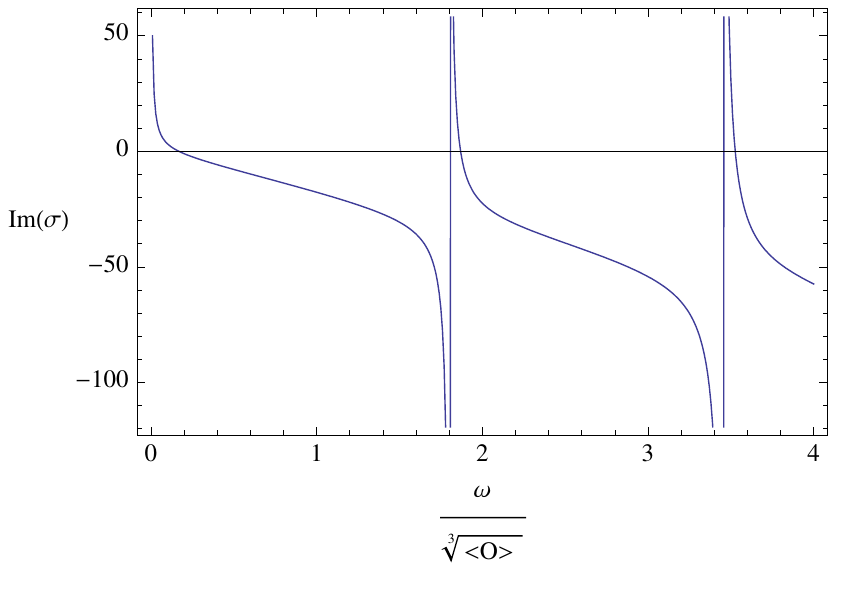}
\caption{ Here we plot the behavior of imaginary part of the conductivity , $\sigma_{yy}$,
for various values of $\omega$ when the expectation value of the dual operator is non-zero.
Note that such a pattern repeated periodically.} \label{fig3}
%\vspace{0.5cm}
\end{center}
\end{figure}

In the figure \ref{fig3} we have plotted the behavior of the conductivity in terms of the
energy when the condensation is non-zero. At $\omega\rightarrow 0$ we find a 
pole showing that we have 
an infinite conductivity as expected for the superconductor phase. Note that the pole structure
shown in figure \ref{fig3} is repeated periodically.  This is due to an infinite tower of 
vector modes.

So far we have seen that the model based on 
the action \eqref{action} exhibits a p-wave insulator/superconductor phase transition
as we vary the chemical potential. On the other hand as we have already mentioned in 
the introduction  the equations of motion
\eqref{eom} admit another analytic solution; the  RN black hole solution. Therefore
one may expect that  the AdS solitonic solution \eqref{metric} could 
decay to an AdS black hole leading 
to another phase transition in the model. This is indeed the  confinement/deconfinement first order phase transition considered in\cite{Witten:1998zw}. In our context from 
three dimensional theory point of view this corresponds to an insulator/conductor 
phase transition. It is then natural to look for another conductor/superconductor phase transition in the model.

We note, however, that to explore the whole picture of the phase structure of the model one 
has to consider the back reactions of the 
gauge field on the metric as well. In other words we will have to  go beyond the probe limit. 
This is, in fact, what we will do in the next section\footnote{The complete phase
diagrams for s-wave holographic insulator/superconductor system has been studied 
\cite{Horowitz:2010jq} where the authors have considered the back reactions of 
the gauge and scalar fields.}.

As a final remark, before going to the back reactions analysis,  
it is an instructive exercise to study the insulator/superconductor phase transition in the
presence of a constant DC current (see for example \cite{Arean:2010xd}).  
To do so we will consider the following ansatz for the
gauge field
\be
A=\phi(r)\sigma^3 dt+\psi(r)\sigma^1 dx+A(r)\sigma^3 dy.
\ee
Note that although we are still working in the probe limit where the back reactions of the gauge
field on the metric are negligible, all components of the gauge fields are treating in the same 
footing. This means that one should  take in to account the effects of  $A_y$ component
on the other components of the gauge field. 
Indeed it turns out that the equation of motion for $\psi$  is modified as follows
\be
 \psi''+\left(\frac{3}{r}+\frac{g'}{g}\right)\psi'+\frac{\phi^2-A^2}{r^4 g}\psi=0,
\ee
 while  for  $\phi$  it remains unchanged. Finally the last equation is given by \eqref{eqA} with $\omega=0$. 

\begin{figure}
\begin{center}
%%\vspace{1cm}
\includegraphics[height=5cm, width=8cm]{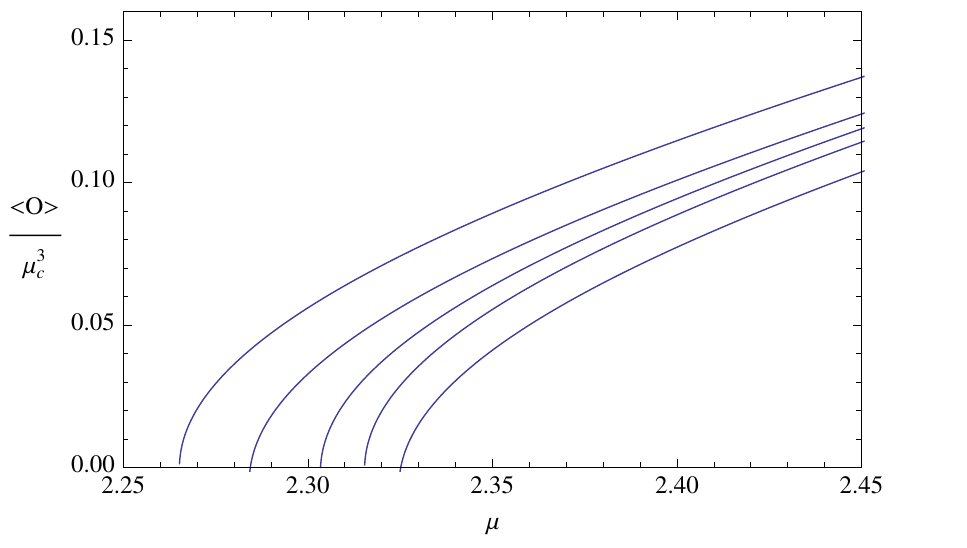}
\caption{ The behavior of the expectation value of the dual operator in terms of the chemical potential in the present  of a constant background field $A_0$. The most left curve is for the case when $A_0=0$, the others 
appear as we increase the field. Therefore as we turn on the background field, 
the condensation is harder to be formed.  } \label{fig4}
%\vspace{0.5cm}
\end{center}
\end{figure}

It is then the aim to solve these equations to find the behavior of the system. Actually 
we find that the model is again unstable against developing  hair leading to the symmetry breaking when the chemical potential is bigger than a critical value. We note, however, that in the present  case the critical value of the chemical increases as we increase
the background DC current (see figure \ref{fig4}). 
In other words having a  non-zero background field makes it 
harder to have insulator/superconductor phase transition.  

\section{Beyond probe limit}

So far we have been considering  the model in the probe limit where the back reactions of the gauge field were negligible.  The aim of this section is to study the effects of the gauge field on the background metric. We, note, that  back reacted solutions  of the 
five dimensional gravity  \eqref{action}  have been studied in \cite{Ammon:2009xh}
(see also \cite{Manvelyan:2008sv})
where the authors have numerically constructed asymptotically AdS charged black holes 
of the model. It was also shown that, for sufficiently low temperature, the model develops
 vector hair. Generically this corresponds to a second order phase transition in the dual field theory. The aim of this section is to further study the back reacted solutions of the theory
given by the action \eqref{action}.

\subsection{AdS soliton}

In this subsection we will study the back reactions of the gauge field on the AdS
solitonic solution considered in the previous section. To proceed,  we start with the following ansatz for the metric and gauge field
\bea\label{ansatz}
ds^2&=&\frac{dr^2}{g(r)}+r^2\left(-f(r)dt^2+h(r)dx^2+dy^2\right)+ g(r)e^{-\chi(r)}d\eta^2,\cr
A&=&\phi(r)\sigma^3 dt+\psi(r)\sigma^1 dx.
\eea
Using this ansatz the Einstein equations of motion read
\bea
&&\frac{f''}{f}+\frac{f'}{f}\left(\frac{3}{r}-\frac{f'}{2f}+\frac{h'}{2h}+\frac{g'}{g}
-\frac{\chi'}{2}\right)=\frac{{2\phi'}^2}{r^2f}+\frac{2q^2\psi^2\phi^2}{r^4fgh},
\cr &&\cr
&&\frac{h''}{h}+\frac{h'}{h}\left(\frac{3}{r}+\frac{f'}{2f}-\frac{h'}{2h}+\frac{g'}{g}
-\frac{\chi'}{2}\right)=-\frac{{2\psi'}^2}{r^2h}+\frac{2q^2\psi^2\phi^2}{r^4fgh},
\cr &&\cr
&&\chi''+\chi'\left(\frac{2}{r}-\frac{3g'}{2g}-\frac{\chi'}{2}\right)-\frac{1}{gr^2}(r^2g-r^4)''
=-\frac{{\psi'}^2}{r^2h}+\frac{{\phi'}^2 }{r^2f}+\frac{3q^2\psi^2\phi^2}{r^4fgh},\cr &&\cr
&&\frac{f'h'}{fh}+\frac{(fh)'}{fh}\left(\frac{1}{r}+\frac{g'}{g}-\chi'\right)-\frac{3\chi'}{r}
=\frac{4q^2\psi^2\phi^2}{r^4fgh},
\eea
while from the equations of motion  of gauge field one finds
\bea
\phi''+\phi' \left(\frac{1}{r}-\frac{f'}{2f}+\frac{h'}{2h}+\frac{g'}{g}-\frac{\chi'}{2}\right)-\frac{q^2\psi^2\phi}{r^2fg}&=&0,\cr &&\cr
\psi''+\psi' \left(\frac{1}{r}+\frac{f'}{2f}-\frac{h'}{2h}+\frac{g'}{g}-\frac{\chi'}{2}\right)+\frac{q^2\phi^2\psi}{r^2hg}&=&0.
\eea
Now the aim is to solve the above equations though in general it requires  numerical
analysis. It is evident that the AdS solitonic solution considered in the previous section 
is, indeed, a solution of the above equation for $\psi=\phi=0$.

Near the boundary where $r\rightarrow \infty$ the behavior of the gauge field is as follows
\be
\phi=\mu-\frac{\rho}{r^2},\;\;\;\;\;\;\;\;\;\;\psi=\psi_0+\frac{\psi_1}{r^2}.
\ee
Of course we set $\psi_0=0$ to avoid having source  for the dual operator in  the field theory. 
The main  goal of this subsection is to find the behavior of  $\psi_1$ as a function of
 the chemical potential
$\mu$ for given electric charge $q$. For this purpose we utilize the numerical method using 
Mathematica. Essentially the procedure is very similar to what we have done in the 
previous section, though here we have to deal with more equations.

To proceed we assume that at the tip, $r=r_0$, the function $g$ goes to zero, {\rm i.e.}
$g(r_0)=0$. On the other hand one may set $r_0=1$ by making use of an evident scaling
symmetry of the equations
\be
r\rightarrow \lambda r,\;\;\;\;\;(t,x,y,\eta)\rightarrow \lambda^{-1}(t,x,y,\eta),\;\;\;\;\;
g\rightarrow \lambda^2 g,\;\;\;\;\;(\phi,\psi)\rightarrow \lambda (\psi,\psi).
\ee
Therefore by considering a series solution near the tip we are left with five parameters
given  $f(1),h(1), \chi(1)$ and $\phi(1),\psi(1)$. We note, however, that 
since the equations are symmetric under  
a constant shift in $\chi$, we may set $\chi=0$ at tip. On the other hand 
due to  the scaling symmetries 
\be 
f\rightarrow a^2 f,\;\;\;\;\;\;\;\;\;\;\;\phi\rightarrow a\phi,
\ee
 and 
\be 
h\rightarrow b^2 h,\;\;\;\;\;\;\;\;\;\;\;\psi\rightarrow b\psi,
\ee
one may  fix $f(1)$ and $h(1)$ by setting, for example,
$f=h=1$ at the boundary. Altogether we are left with two parameters
 $\psi(1)$ and $q$. Therefore for a given $q$ we have a one parameter family of solutions
parametrized by the value of $\psi$ at the tip.

To summarize we can solve the equations by treating $\phi(1)$ as the shooting parameter for a fixed $q$ and given $\psi(1)$. Then we can scan the moduli space of parameters by changing 
$q$. Doing so we find the following picture. 

 For generic $q$ the behaviors of $\psi$ and $\rho$ 
as the functions of the chemical potential are indeed the same as those we had in the 
probe limit (see figures \ref{fig1}).  More precisely  there is a critical chemical potential above which the system becomes unstable to develop vector hair leading to a second order phase transition in the dual field theory.

We, note, however that in the present case the critical value of the chemical potential depends on the value of $q$. 
Indeed as we increase  $q$ the value of the chemical potential, where the phase transition 
occurs, decreases. An interesting phenomena we encounter is that there is a
critical value for $q$ ($q\approx 1.94$)  below which the system exhibits qualitatively 
different behaviors.
This phenomena is also seen in the AdS charged black hole which we will discuss in the 
next subsection.  
 Actually the situation is very similar to the  s-wave holographic insulator/ superconductor system studied in \cite{Horowitz:2010jq}. We will back to this point latter

\subsection{AdS charged black hole}

The model given by the action \eqref{action} admits AdS charged black holes when the
 back reactions of the gauge field are taken into account.  These solutions have 
been numerically studied
in \cite{Ammon:2009xh} where the authors have also shown that the solutions
are unstable to develop a vector hair for sufficiently low temperature. 
 In this subsection, for completeness of our study,  we will rederive the results of  \cite{Ammon:2009xh}. Of course 
our ansatz are slightly different from that in  \cite{Ammon:2009xh},  though our
final results are similar them,  up to a normalization.

Motivated by \cite{Basu:2009vv} we consider the following ansatz 
\bea\label{ansatz2}
ds^2&=&\frac{dr^2}{g(r)}+r^2\left(h(r)dx^2+dy^2+dz^2\right)-g(r)e^{-\chi(r)}dt^2,\cr
A&=&\phi(r)\sigma^3 dt+\psi(r)\sigma^1 dx.
\eea
Plugging the above ansatz into the equations of motion \eqref{eom} for thr metric 
  one finds 
\bea\label{eom2}
&&-\frac{g'}{g}\left(\frac{3}{2r}+\frac{h'}{2h}\right)-\left(\frac{3}{r^2}
+\frac{4h'}{rh}+\frac{h''}{h}\right)+\frac{6}{g}
=\frac{e^\chi}{4g}\phi'^2+\frac{1}{4r^2h^2}\psi'^2+\frac{q^2e^\chi}{4r^2g^2h^2}\phi^2\psi^2,
\cr &&\cr
 &&\frac{h''}{h}+\frac{h'}{h}\left(\frac{3}{r}+\frac{g'}{g}-\frac{\chi'}{2}\right)=-\frac{\psi'^2}{2r^2h^2}+\frac{q^2e^\chi}{2r^2g^2h^2}\phi^2\psi^2,
\cr &&\cr
&& \frac{h'}{h}\left(\frac{1}{r}+\frac{g'}{g}-\chi'\right)-\frac{3\chi'}{2r}=\frac{q^2e^\chi}{r^2g^2h^2}\phi^2\psi^2,
\eea
while for the gauge field we get
\bea
\phi''+\phi'\left(\frac{3}{r}+\frac{\chi'}{2}+\frac{h'}{h}\right)-\frac{q^2\psi^2}{r^2gh^2}\phi=0,
\;\;\;\;\;\psi''+\psi'\left(\frac{1}{r}+\frac{g'}{g}-\frac{h'}{h}-\frac{\chi'}{2}\right)+\frac{e^\chi
q^2\phi^2}{g^2}\psi=0.\nonumber
\eea
For finite $q$ where the back reactions of the gauge field are important the model admits
AdS charge black holes. In particular when we set $\psi=0$ the equations admit
an analytic solution which is indeed  the Reissner-Nordstrom AdS black hole given
by \eqref{RN}.

For generic values of the gauge field, beside the RN black hole, there is no other well known
 analytic solution, though 
asymptotically AdS charged black holes have been numerically constructed in \cite{Ammon:2009xh} (see also \cite{Manvelyan:2008sv}). Using the same procedure as 
that in the previous sections we may solve the equations numerically. To proceed we note that 
for large $r$ as we approach the boundary the behavior of the gauge field is as follows
\be
\phi=\mu-\frac{\rho}{r^2},\;\;\;\;\;\;\;\;\;\psi=\psi_0+\frac{\psi_1}{r^2}.
\ee
Again we will set $\psi_0=0$.
On the other hand near the horizon where $g$ vanishes we assume $\phi$ is zero too. 
The horizon may be set to be located at $r=1$.  By making use of the time reparamerization we can set $\chi$ to zero at the horizon. On the other hand utilizing the following symmetries
\be
r\rightarrow \lambda r,\;\;\;\;g\rightarrow \lambda^2 g,\;\;\;\;\;(\phi,\psi)\rightarrow \lambda
(\phi,\psi),
\ee
and 
\be
h\rightarrow \xi h,\;\;\;\;\;\psi\rightarrow \xi \psi,
\ee
one may fix the values of  $g$ and $h$ at the horizon by setting , for example, 
$g=r^2$ and $h=1$ at the boundary where we recove the AdS solution.

Taking into account that the value of $\phi$ at the horizon is treated as the shooting parameter  we are left with two parameters given by $q$ and $\psi$ at the horizon. Therefore  for a 
fixed $q$ we get one parameter family of solutions parametrized by the value of $\psi$ at the horizon.
 
\begin{figure}
\begin{center}
%\vspace{1cm}
\includegraphics[height=5cm, width=8cm]{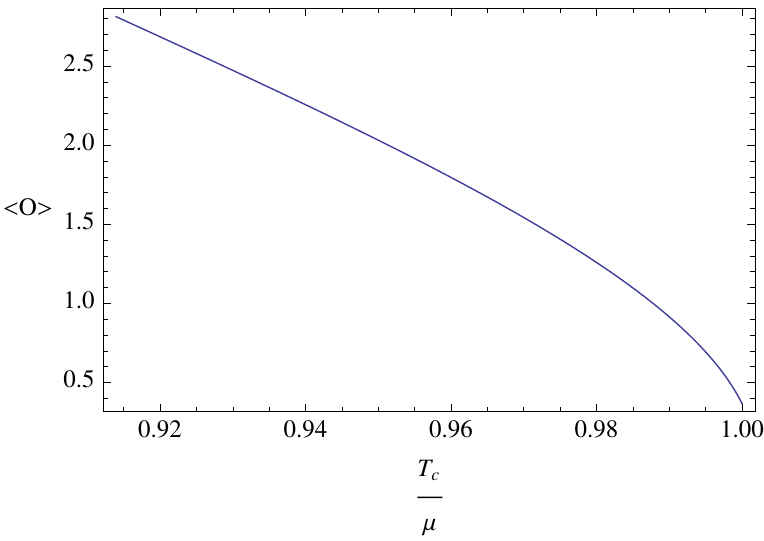}
\caption{  The behavior of the expectation value in terms of temperature for $q=3.5$ where 
we get $T_c=0.226$.} \label{fig6}
%\vspace{0.5cm}
\end{center}
\end{figure}

For generic $q$ one finds that   the solution is unstable to develop a vector hair for sufficiently low temperature where  the $U(1)$ gauge symmetry is also broken.  This corresponds to a second order conductor/superconductor phase transition from field theory point of view (see figure \ref{fig6})\cite{Ammon:2009xh}. 

We note that the critical temperature depends on the value of $q$. Indeed as we increase $q$ the 
critical temperature increases too. It is worth mentioning that there is a critical value for 
$q$ where the system shows qualitatively different behaviors.  In the normalization of our 
 numerical computations it happens at $q\approx 1.94$. This is exactly the behavior obtained in \cite{Ammon:2009xh} where 
the authors have found that there is a critical $q$ where the transition becomes first
order. This has to be compared with behavior we have found in the previous section. In the
following section we will discuss about this point.

\subsection{Phase structure}

In this subsection we would like to explore the phase structure of  the model using the
results we have found so far.  To proceed we note that the equations of motion support two
distinctive solutions; AdS soliton and AdS charged black hole. In each case the solution
becomes unstable to develop a vector hair as we change the parameters of the model.
In particular we have seen that when we change the chemical potential there is a 
critical point above which the AdS soliton develops a vector hair. From field theory
point of view this  phenomena 
corresponds to the insulator/superconductor second order phase transition. 
On the other hand 
when we have a charged black hole the solution is unstable to generate a vector hair for 
sufficiently low temperature which is indeed the holographic realization of the second order
conductor/superconductor phase transition.

\begin{figure}
\begin{center}
%\vspace{1cm}
\includegraphics[height=3.5cm, width=5cm]{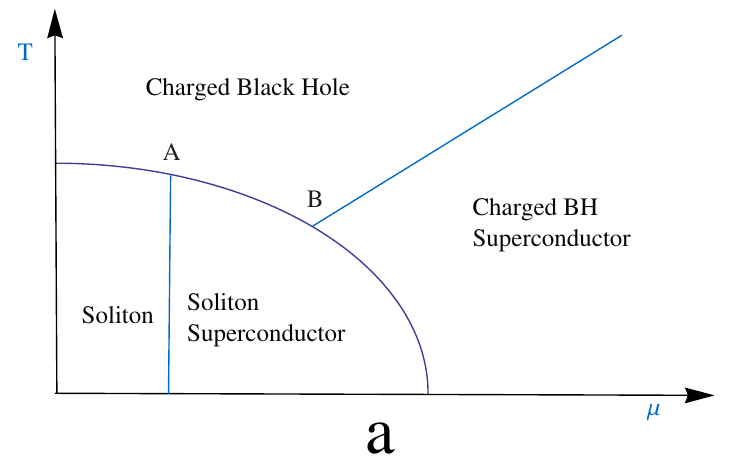}
\includegraphics[height=3.5cm, width=5cm]{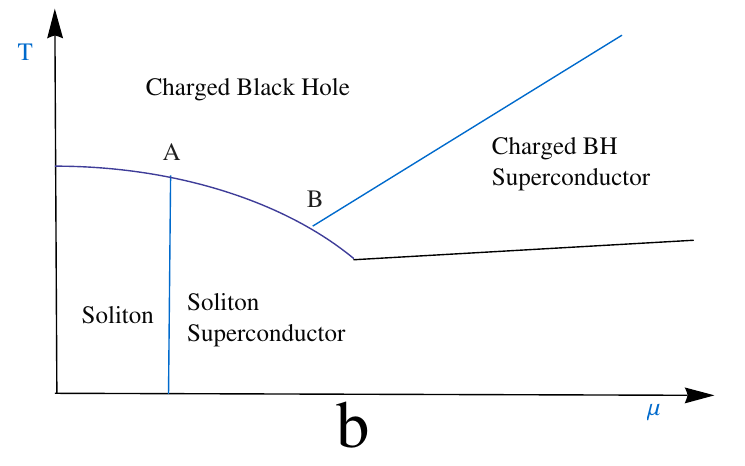}
\includegraphics[height=3.5cm, width=5cm]{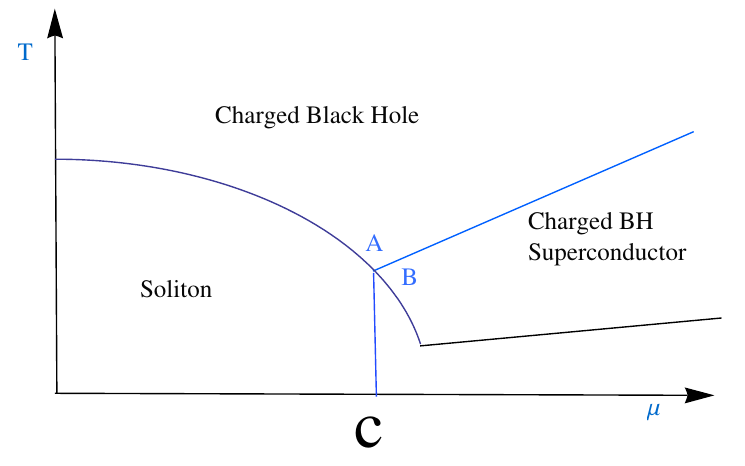}
\includegraphics[height=3.5cm, width=5cm]{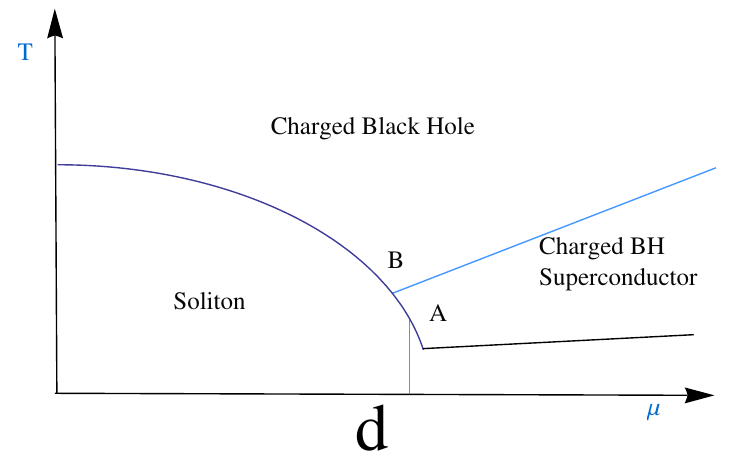}
\caption{The qualitative phase structure of the model. As we change $q$ the positions of 
points A and B are changed and therefore the shape of different regions get modified. There is a critical point where A and B are on top of each other. For sufficiently small $q$ there is a possibility to have a first order phase transition where a superconductor becomes and insulator as we decrease temperature  as shown in the left picture.  The figure ``a''  is the phase structure
in the probe limit, while the others are with back reactions. } \label{gen}
%\vspace{0.5cm}
\end{center}
\end{figure}

On top of these we note that using the Euclidean solitonic solution one may associate a
temperature to the solution which could be any value for a given chemical potential. Therefore
as we draw the phase structure in the $T-\mu$ space the curve which separates insulator phase from the superconductor phase should be a line parallel to the $T$ axes.  On the other hand when the period of the Euclidean time of the AdS soliton becomes the same as that in the AdS 
charged black hole the free energy of the system shows that the favored solution is  
the AdS charged black hole  and thus we have a first order phase transition 
from AdS soliton to AdS charged black hole\cite{Witten:1998zw}.
So altogether we get four different phases as shown in figure \ref{gen}-a. 

 It should be mentioned that in order 
to find the above picture for the phase structure of the model the results we have  gathered  from 
probe limit considerations were sufficient.  We note, however, that in order to fully explore  different phases of the model
in more detail one needs to go
beyond the probe limit. In particular taking into account the back reactions we observe two new
features in the phase diagram of the model.

The first observation is that for large chemical potentail as we decrease temperature the favored 
phase is soliton superconductor. In other words the phase diagram gets modifed as figure
\ref{gen}-b. In particular  we cannot have a phase describing hairy charged black hole at zero
temperature which could have been the case if the phase diagram had been given by figure
\ref{gen}-a. We will back to this point in the next section. It is important to note that the figure \ref{gen} should be treated as a qualitative
description and the interfaces between different phases have been ploted schematically.
To explore the precise structure of the phase diagram higher numerical precision is needed,
specially for small $q$. We hope to back to this point in our future work.

On the other hand  there are two special points, named by A and B in  figure
\ref{gen} which in order  to understand their physical significant, it requires to study the system beyond the probe limit. 

\begin{figure}
\begin{center}
%\vspace{1cm}
\includegraphics[height=4cm, width=6cm]{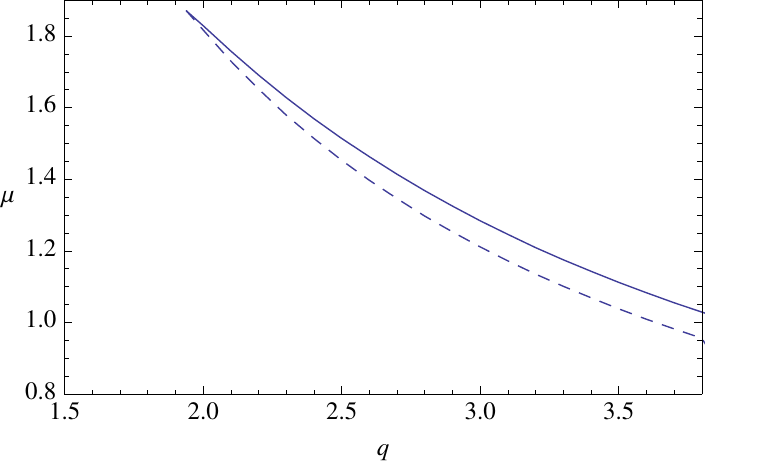}
\caption{The behavior of $\mu_A$ in terms $q$ is shown by the dashed line, while the solid line
is for $\mu_B$. Theses two lines meet each other at $q\approx 1.94$ where  $\mu_B\approx\mu_A\approx 1.87$.  } \label{AB}
%\vspace{0.5cm}
\end{center}
\end{figure}

Actually our numerical computations show that the positions of these two points labeled by 
$\mu_A$ and $\mu_B$ change as we are  changeing  $q$. More precisely, when we decrease $q$, $\mu_A$ and $\mu_B$ increase as well and eventually  they meat each other as we approach the critical point $q=1.94$ where one gets 
$\mu_B\approx\mu_A\approx 1.87$  (see figure \ref{AB}) .

As the result  when we change the parameter $q$, different regions in the phase diagram of the model get modified and in particular we reach a situation   where two points $A$ and $B$ are on top of each other as shown in  figure \ref{gen}-c.

This is indeed the critical point we have found in our numerical calculations in the previous
subsections where the model shows qualitatively different behavior. This is also the situation 
which has  been observed in \cite{Ammon:2009xh} ( and also in \cite{Horowitz:2010jq}
for s-wave) where it was shown that  the phase transition becomes first order at the critical point. 

As we further decrease $q$ we observe that  while  both $\mu_A$ and $\mu_B$ increase, the  point A passes 
though the point  B  (see figure \ref{gen}-d). In this case we 
encounter a  new phase transition. Actually as we decrease temperature there is a range of
$\mu$ bewteen which  the superconductor becomes an insulator via a first order phase transition. This 
phase transition has been also seen in the s-wave consideration in \cite{Horowitz:2010jq}.
Going further it seems that the phase where we have soliton superconductor becomes smaller
and smaller and eventually disappears from the phase diagram, though due to the uncertainty of
our numerical results, we have not been able to explore the situation exactly.  In particular
for low temperaute (for small enough $q$, i.e. $q\approx 0.86$) the numerical solution develops a singularity and 
one has to study $T\rightarrow 0$ limit of hairy charged black hole more carefully. This is indeed
what we will do in the next section.

\section{Zero temperature limit}

So far  we have been considering the phase transition 
between AdS soliton and AdS charged black hole with non-zero temperature. 
It is then natural to pose the equation what happens 
when we send the temperature of the AdS charged black hole to zero?\footnote{Note that  
the insulator/superconductor phase transition we have considered in the previous
 section was also
occurred at zero temperature. We note, however, that this has not to be compared with the present case. This is because in the previous case to get the solitonic solution we have put 
anti-periodic boundary condition for 
the fermions along the compact direction which is not thr case here.}. In general for a charged black hole sending temperature to zero we will end up with an extremal black hole whose near horizon
geometry  develops an $AdS_2$ throat with non-zero entropy. Therefore if we would like to 
study holographic superconductors at zero temperature the extremal black hole cannot provide the gravity dual descriptions. In fact to get an eligible background one must have a 
geometry with 
zero size  horizon ensuring that the ground state is a single state (entropy is zero). 

Indeed zero  temperature limit of  s-wave holographic  superconductors 
in three dimensions has been
 studied in \cite{Horowitz:2009ij} (see also\cite{{Gubser:2008wz},
{Gubser:2009cg},{Gauntlett:2009bh},{Konoplya:2009hv}}) where it was shown that the corresponding 
hairy solution develops an $AdS_4$ geometry at the core of the space time. More precisely they
have found  numerical solutions which interpolate between   hairy $AdS_4$ and 
 $AdS_4$ geometries. The Zero  temperature limit of  the holographic p-wave superconductors 
in three dimensions has also been studied in \cite{Basu:2009vv}.

To complete our discussions of  p-wave holographic superconductors in four 
dimensions, in this section  we will consider  zero  temperature limit of  the
four dimensional p-wave holographic superconductor.
Actually to do this one needs to solve the equations of motion \eqref{eom2} with a
particular boundary condition ensuring that the resultant geometry would have
zero size horizon. To proceed we consider the following behaviors for the parameters of the 
ansatz \eqref{eom2} in  $r\rightarrow0^+$ limit.
\be
\phi\sim \phi_{0}(r),\;\;\;\;\psi\sim \psi_{0}-\psi_{1}(r),\;\;\;\;
\chi\sim \chi_{0}-\chi_{1}(r),\;\;\;\; g\sim
r^2+g_{1}(r),\;\;\;\;h\thicksim h_{0}+h_{1}(r),
\ee
with the assumption that $\phi_0,\psi_1,\chi_1,g_1$ and $h_1$ go to zero
sufficiently fast in the limit of   $r\rightarrow0^+$. Plugging these behaviors into the
corresponding  equations of motion, at leading order, one finds
\bea
&&\phi=\phi_{0}\frac{e^{-\frac{\alpha}{r}}}{\sqrt{r}},\;\;\;\;
\chi=\chi_{0}-\frac{e^{\chi_{0}}\alpha
\phi_{0}^2}{6r^2}e^{-\frac{2\alpha}{r}},\;\;\;\;g=r^2-\frac{e^{\chi_{0}}\alpha
\phi_{0}^2}{6r^2}e^{-\frac{2\alpha}{r}},\cr &&\cr&&\cr
&& \psi=\psi_{0}\left(1-\frac{e^{\chi_{0}}q^2\phi_{0}^2}{4r\alpha^2}
e^{-\frac{2\alpha}{r}}\right),\;\;\;\;\;\;\;\;h=h_{0}\left(1+\frac{e^{\chi_{0}}\phi_{0}^2}{8r}e^{-\frac{2\alpha}{r}}\right).
\eea
where $\alpha=q\psi_0/h_0$.
On the other hand for large $r$ one impose the following asymptotic conditions for the  gauge field components
\be
\phi= \mu-\frac{\rho}{r^2},\;\;\;\;\;\;\;\;\;\psi=\psi_0+\frac{\psi_1}{r^2}.
\ee 
Now the aim is to solve numerically the  equations of motion with the above boundary conditions. To do so, using
the symmetries of the model we can set $\phi_0=h_0=1$ and $\chi_0=0$.
On the other hand we would like to have a solution which is asymptotically $AdS_5$ without
hair.  Therefore we are interested in the case where $\psi$ vanishes asymptotically. This 
can be done by imposing a relation between  $\alpha$ and  $q$.  For example for $3<q<7$ one finds $\alpha=0.8+0.49 q$. 

\begin{figure}
\begin{center}
%\vspace{1cm}
\includegraphics[height=3.5cm, width=5cm]{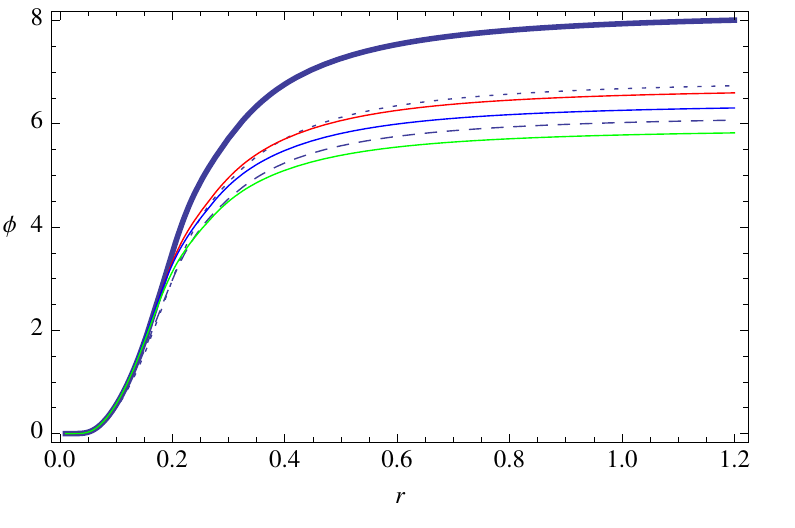}
\includegraphics[height=3.6cm, width=5cm]{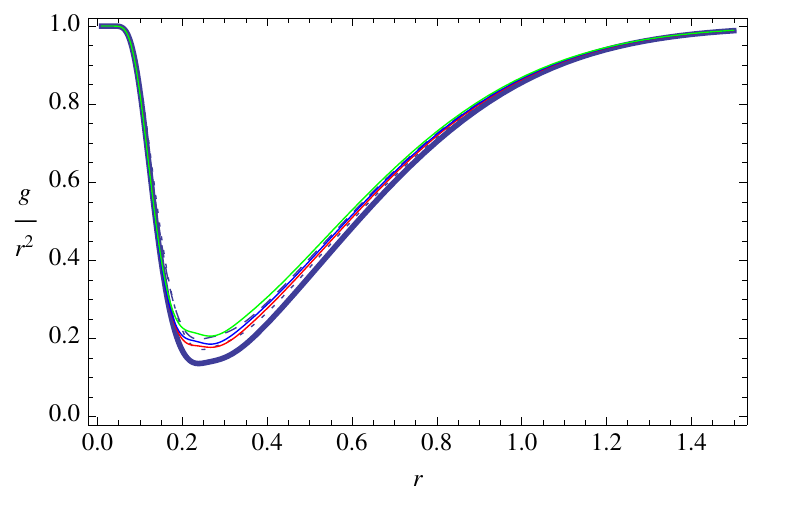}
\includegraphics[height=3.5cm, width=5cm]{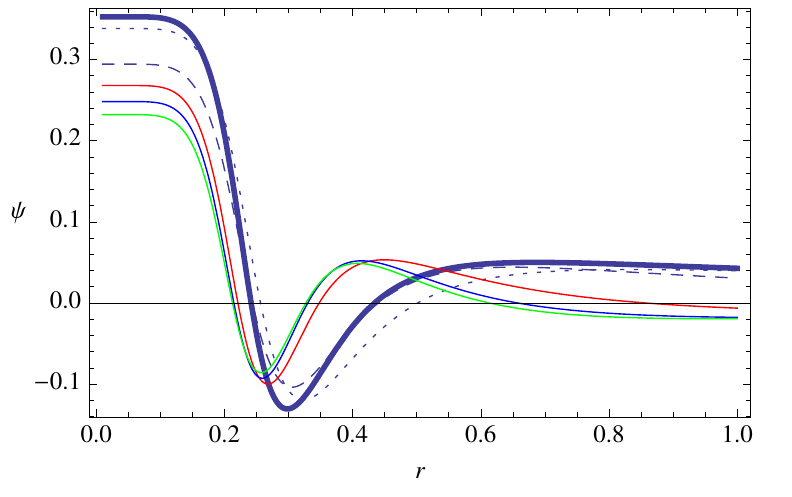}
\includegraphics[height=3.5cm, width=5cm]{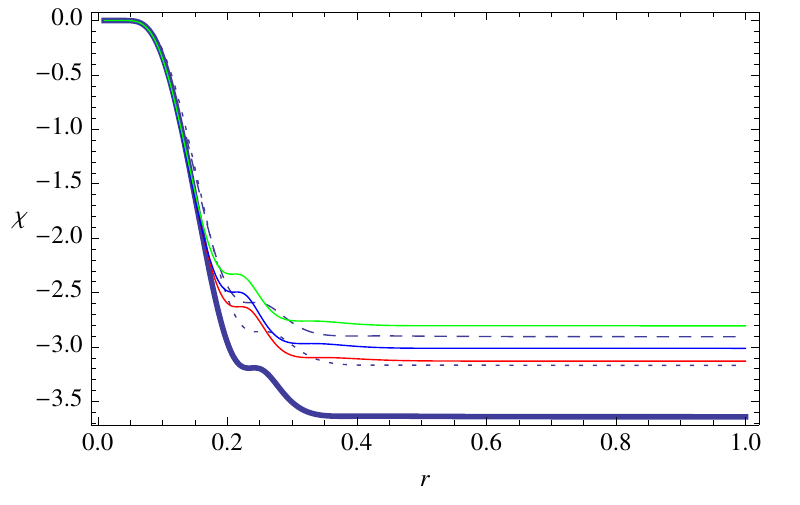}
\includegraphics[height=3.5cm, width=5cm]{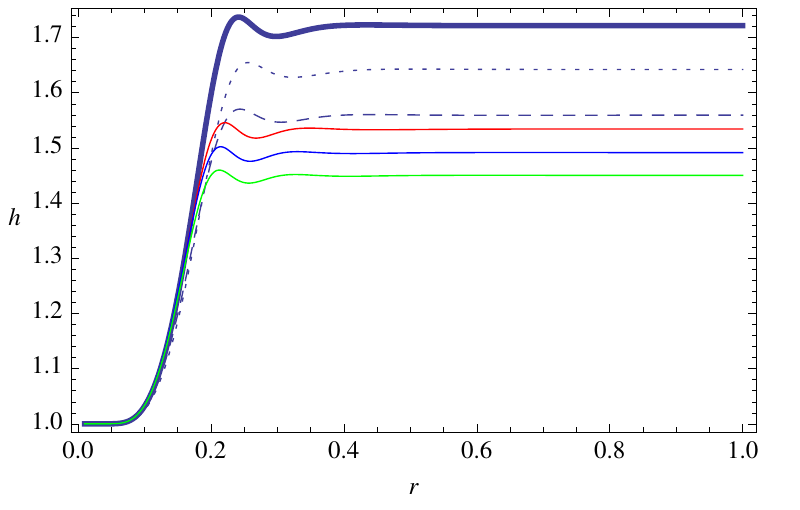}
\caption{The behavior of $\phi,g/r^2,\psi,\chi$ and $h$ as  functions of $r$.  The function is plotted for $q=0.93,1,1.12,1.22,1.32,1.41$ which have been  shown by thick, dashed,dotted,  red, blue and green curves respectively.} \label{zero}
%\vspace{0.5cm}
\end{center}
\end{figure}

It is straightforward to solve the equations numerically using Mathematica. The 
final results are shown in figure \ref{zero}.
As we see the solutions approach $AdS_5$ geometries at IR with non-zero hair while
they are asymptotically $AdS_5$ without hair. Therefore from field theory
point of view in IR we get  an emergent CFT broken by the expectation value of the operator
that is dual to $\psi$.

It is worth noting  that as we approach $q\rightarrow \sqrt{3}/2$ limit we lose the 
accuracy of our numerical solutions. This is the limit where we recover the extremal
limit of the charged AdS black hole. Therefore  effectively we will have to  solve the equations on 
the $AdS_2$ geometry and $\sqrt{3}/2$ is, indeed, the Breitenlohner-Freedman bound for mass
in $AdS_2$ space time.

Having found hairy solutions it is natural to compute  the conductivity for the model.
To study the conductivity we consider small fluctuations for the metric and $A_y$ component
of the gauge field as follows
\be
g_{ty}=\epsilon
f(r)e^{-i\omega t},\;\;\;\;\;\;\;\;\;\;A^3_{y}=\epsilon \varphi(r)e^{-i\omega t}.
\ee
where $\epsilon$  is a small number controls the effects of the perturbations.
As far as the equations of motion for gauge field  and metric are concerned, since we 
are working in the probe limit, the effects of the perturbations on the equations of
motion of the other components  are negligible and thus one can still use the equations
\eqref{eom2}.  On the other hand  for the perturbations, at first order in $\epsilon$, the equations of motion are given by
\bea
&&\varphi''+\varphi'\left(\frac{1}{r}+\frac{g'}{g}-\frac{\chi'}{2}+\frac{h'}{h}\right)+\varphi
\left(\frac{e^\chi\omega^2}{g^2}-\frac{q^2\psi^2}{r^2gh^2}\right)
+\frac{e^\chi f}{g}\bigg[\phi''
+\phi'\left(\frac{1}{r}+\frac{f'}{f}+\frac{\chi'}{2}+\frac{h'}{h}\right)\cr &&\cr &&
\;\;\;\;\;\;\;\;\;\;\;\;\;\;
-\frac{q^2\psi^2}{r^2gh^2}\phi\bigg]=0,
\;\;\;\;\;\;\;\;\;\;\;\;\;\;\;\;\;\;r^2f'-2rf+r^2\varphi\phi'=0.
\eea
By making use  of the equation of motion for $\phi$ , \eqref{eom2}, one arrives at 
\be\label{var}
\varphi''+\varphi'\left(\frac{1}{r}+\frac{g'}{g}-\frac{\chi'}{2}+\frac{h'}{h}\right)
+\varphi\left(\frac{e^\chi\omega^2}{g^2}-\frac{q^2\psi^2}{r^2gh^2}-e^\chi\frac{\phi'^2}{g}\right)=0.
\ee
To find the conductivity one needs to find the asymptotic behavior of 
gauge field $\varphi$.  In fact for large $r$ where we find 
$\varphi=\varphi_0+\frac{\varphi_1}{r^2}$ the conductivity is given by
\be\label{sigma}
\sigma=\frac{-i \varphi_1}{\omega\varphi_0}.
\ee
On the other hand at $r\rightarrow 0$, using the results of the last section 
 and imposing in coming boundary condition at $r\rightarrow 0$ we find
\be
\varphi(r\rightarrow 0)=\tilde{\varphi}_0 \frac{e^{i\frac{\tilde{\omega}}{r}}}{\sqrt{r}}
\ee
where $\tilde\omega=\sqrt{\omega^2-\alpha^2}$.

Taking into account that  the solutions we have found for the parameters of the anstaz
\eqref{ansatz2} are real and using the equation \eqref{var} together with its complex
conjugate we can show that the following expression is a constant of motion in the sense 
that it is $r$ independent
\be
{\rm Im} \left(rhge^{-\chi/2}\varphi^*{\varphi}'\right).
\ee
Therefore its value at the horizon is the same  as that at the boundary.  Evaluating the 
above expression at the boundary and near $r=0$ one finds 
$2 {\rm Im}(\varphi_0^*\varphi_1)=i|\tilde\varphi_0|^2 {\rm Re}(\tilde\omega)$. Plugging 
this result into \eqref{sigma}  we arrive at
\be
{\rm Re}(\sigma)=\frac{{\rm Re}(\tilde\omega)
}{2\omega}\frac{|\tilde\varphi_0|^2}{|\varphi_0|^2},
\ee
that is zero for $\omega<\alpha$ leading to a gap. Indeed, this is  exactly (up to a factor of 1/2)
 the same expression which
found in \cite{Basu:2009vv} for three dimensional model.

\section{Conclusions}

In this paper we have studied holographic insulator/superconductor by making use of a five 
dimensional Einstein-Yang-Mills theory with an $SU(2)$ gauge field.
The equations of motion supports both AdS soliton as well as AdS charge black hole
solutions. Both solutions may become unstable to forming a vector hair as we change 
the parameters of the solutions, including temperature and chemical potential.

Therefore altogether the model exhibits four different phases. From gravity point of view
they correspond to AdS soliton, hairy AdS soliton, AdS charged black hole and hairy AdS 
charged black hole and from field theory point of view they are insulator, superconductor,
 conductor and another superconductor, respectively. As we see there are two phases 
where the system  is in the superconductor phase. 
 Actually  these two phases can be distinguished by the behavior of the conductivity \cite{Horowitz:2010jq}; while in the solitonic case
the real part of the conductivity has a series of delta functions, in the charged black hole 
case we just  have  a gap at low frequency.

Using the probe limit we have been able to study  qualitatively the general features of the phase structure of the system, though to explore the phase diagram in more detail it was curtail 
to take into account the back reactions. Doing so, we have found that the model has rather
 a rich phase structure. In particular when $q<1.94$ there is a range of $\mu$ between which
the system can show a first order phase transition from superconductor to insulator as we decrease temperature. Indeed 
this is a new phenomena in  which has been first
observed in the context of  s-wave insulator/superconductor
transition in \cite{Horowitz:2010jq}. It would be interesting to understand
this phase transition better.

We have also studied zero-temperature limit of four dimensional p-wave holographic
superconductor. We have numerically solved the corresponding equations of motion 
have shown  that equations of motion admit smooth solutions which interpolate
between  hairy $AdS_5$ solution at IR and $AdS_5$ solutions without  hair at UV.
Having found a hairy $AdS_5$ solution at IR shows that there is an  emergent CFT at IR such
that  the conformal symmetry is broken 
by the expectation value of the dual operator.  
Evaluating the conductivity for the solutions one finds that the real part of the 
conductivity vanishes for $\omega<\alpha$ leading to a gap in the model.

\section*{Note:} 
All results of the present paper are based on several Mathematica
codes we have written for each part.  We should admit that to develop our Mathematica 
codes the one  prepared by C. P. Herzog which is available at his homepage
(http://wwwphy.princeton.edu/~cpherzog/)  was
illustrative.

\section*{Acknowledgments} 
We would like to thank H. Allahbakhshi, R. Asgari, A. Davody and  A. Mosaffa for 
useful discussions. M. A. would also like
to thank CERN TH-Division where the project started for warm hospitality.
This work is supported by Iran National Science Foundation (INSF).
We would like to thank referee for his/her useful comments.

%--------------------------------------------------------------------

\begin{thebibliography}{39}

%\cite{Maldacena:1997re}
\bibitem{Maldacena:1997re}
  J.~M.~Maldacena,
  ``The large N limit of superconformal field theories and supergravity,''
  Adv.\ Theor.\ Math.\ Phys.\  {\bf 2}, 231 (1998)
  [Int.\ J.\ Theor.\ Phys.\  {\bf 38}, 1113 (1999)]
  [arXiv:hep-th/9711200].
  %%CITATION = IJTPB,38,1113;%%


%\cite{Gubser:2008px},
\bibitem{Gubser:2008px}
  S.~S.~Gubser,
  ``Breaking an Abelian gauge symmetry near a black hole horizon,''
  Phys.\ Rev.\  D {\bf 78}, 065034 (2008)
  [arXiv:0801.2977 [hep-th]].
  %%CITATION = PHRVA,D78,065034;%%

%\cite{Hartnoll:2008vx},{Hartnoll:2008kx}
\bibitem{Hartnoll:2008vx}
  S.~A.~Hartnoll, C.~P.~Herzog and G.~T.~Horowitz,
  ``Building a Holographic Superconductor,''
  Phys.\ Rev.\ Lett.\  {\bf 101}, 031601 (2008)
  [arXiv:0803.3295 [hep-th]].
  %%CITATION = PRLTA,101,031601;%%

%\cite{Hartnoll:2008kx}
\bibitem{Hartnoll:2008kx}
  S.~A.~Hartnoll, C.~P.~Herzog and G.~T.~Horowitz,
  ``Holographic Superconductors,''
  JHEP {\bf 0812}, 015 (2008)
  [arXiv:0810.1563 [hep-th]].
  %%CITATION = JHEPA,0812,015;%%

%\cite{Nishioka:2009zj}
\bibitem{Nishioka:2009zj}
  T.~Nishioka, S.~Ryu and T.~Takayanagi,
  ``Holographic Superconductor/Insulator Transition at Zero Temperature,''
  JHEP {\bf 1003}, 131 (2010)
  [arXiv:0911.0962 [hep-th]].
  %%CITATION = JHEPA,1003,131;%%

%\cite{Witten:1998zw}
\bibitem{Witten:1998zw}
  E.~Witten,
   ``Anti-de Sitter space, thermal phase transition, and confinement in  gauge
  theories,''
  Adv.\ Theor.\ Math.\ Phys.\  {\bf 2}, 505 (1998)
  [arXiv:hep-th/9803131].
  %%CITATION = 00203,2,505;%%

%\cite{Horowitz:2010jq}
\bibitem{Horowitz:2010jq}
  G.~T.~Horowitz and B.~Way,
  ``Complete Phase Diagrams for a Holographic Superconductor/Insulator
  System,''
  arXiv:1007.3714 [hep-th].
  %%CITATION = ARXIV:1007.3714;%%





%\cite{Gubser:2008zu}
\bibitem{Gubser:2008zu}
  S.~S.~Gubser,
  ``Colorful horizons with charge in anti-de Sitter space,''
  Phys.\ Rev.\ Lett.\  {\bf 101}, 191601 (2008)
  [arXiv:0803.3483 [hep-th]].
  %%CITATION = PRLTA,101,191601;%%


%\cite{Gubser:2008wv}{Gubser:2008zu}{Roberts:2008ns}
\bibitem{Gubser:2008wv}
  S.~S.~Gubser and S.~S.~Pufu,
  ``The gravity dual of a p-wave superconductor,''
  JHEP {\bf 0811}, 033 (2008)
  [arXiv:0805.2960 [hep-th]].
  %%CITATION = JHEPA,0811,033;%%

%\cite{Roberts:2008ns}
\bibitem{Roberts:2008ns}
  M.~M.~Roberts and S.~A.~Hartnoll,
  ``Pseudogap and time reversal breaking in a holographic superconductor,''
  JHEP {\bf 0808}, 035 (2008)
  [arXiv:0805.3898 [hep-th]].
  %%CITATION = JHEPA,0808,035;%%




%\cite{Aprile:2010ge}
\bibitem{Aprile:2010ge}
  F.~Aprile, D.~Rodriguez-Gomez and J.~G.~Russo,
  ``p-wave Holographic Superconductors and five-dimensional gauged
  Supergravity,''
  arXiv:1011.2172 [hep-th].
  %%CITATION = ARXIV:1011.2172;%%


%\cite{Gubser:1998bc}
\bibitem{Gubser:1998bc}
 S.~S.~Gubser, I.~R.~Klebanov and A.~M.~Polyakov,
``Gauge theory correlators from non-critical string theory,''
Phys.\ Lett.\  B {\bf 428}, 105 (1998)
  [arXiv:hep-th/9802109].
%%CITATION = PHLTA,B428,105;%%

%\cite{Witten:1998qj}
\bibitem{Witten:1998qj}
  E.~Witten,
  ``Anti-de Sitter space and holography,''
  Adv.\ Theor.\ Math.\ Phys.\  {\bf 2}, 253 (1998)
  [arXiv:hep-th/9802150].
  %%CITATION = 00203,2,253;%%


%\cite{Basu:2008bh}{Ammon:2008fc}
\bibitem{Basu:2008bh}
  P.~Basu, J.~He, A.~Mukherjee and H.~H.~Shieh,
  ``Superconductivity from D3/D7: Holographic Pion Superfluid,''
  JHEP {\bf 0911}, 070 (2009)
  [arXiv:0810.3970 [hep-th]].
  %%CITATION = JHEPA,0911,070;%%

%\cite{Ammon:2008fc}
\bibitem{Ammon:2008fc}
  M.~Ammon, J.~Erdmenger, M.~Kaminski and P.~Kerner,
  ``Superconductivity from gauge/gravity duality with flavor,''
  Phys.\ Lett.\  B {\bf 680} (2009) 516
  [arXiv:0810.2316 [hep-th]].
  %%CITATION = PHLTA,B680,516;%%



%\cite{Arean:2010xd}
\bibitem{Arean:2010xd}
  D.~Arean, M.~Bertolini, J.~Evslin and T.~Prochazka,
  ``On Holographic Superconductors with DC Current,''
  JHEP {\bf 1007}, 060 (2010)
  [arXiv:1003.5661 [hep-th]].
  %%CITATION = JHEPA,1007,060;%%


%\cite{Ammon:2009xh}
\bibitem{Ammon:2009xh}
  M.~Ammon, J.~Erdmenger, V.~Grass, P.~Kerner and A.~O'Bannon,
  ``On Holographic p-wave Superfluids with Back-reaction,''
  Phys.\ Lett.\  B {\bf 686}, 192 (2010)
  [arXiv:0912.3515 [hep-th]].
  %%CITATION = PHLTA,B686,192;%%

%\cite{Manvelyan:2008sv}
\bibitem{Manvelyan:2008sv}
  R.~Manvelyan, E.~Radu and D.~H.~Tchrakian,
  ``New AdS non Abelian black holes with superconducting horizons,''
  Phys.\ Lett.\  B {\bf 677}, 79 (2009)
  [arXiv:0812.3531 [hep-th]].
  %%CITATION = PHLTA,B677,79;%%

%\cite{Horowitz:2009ij}
\bibitem{Horowitz:2009ij}
  G.~T.~Horowitz and M.~M.~Roberts,
  ``Zero Temperature Limit of Holographic Superconductors,''
  JHEP {\bf 0911}, 015 (2009)
  [arXiv:0908.3677 [hep-th]].
  %%CITATION = JHEPA,0911,015;%%

%\cite{Gubser:2008wz}
\bibitem{Gubser:2008wz}
  S.~S.~Gubser and F.~D.~Rocha,
   ``The gravity dual to a quantum critical point with spontaneous symmetry
  breaking,''
  Phys.\ Rev.\ Lett.\  {\bf 102}, 061601 (2009)
  [arXiv:0807.1737 [hep-th]].
  %%CITATION = PRLTA,102,061601;%%

%\cite{Gubser:2009cg}
\bibitem{Gubser:2009cg}
  S.~S.~Gubser and A.~Nellore,
  ``Ground states of holographic superconductors,''
  Phys.\ Rev.\  D {\bf 80}, 105007 (2009)
  [arXiv:0908.1972 [hep-th]].
  %%CITATION = PHRVA,D80,105007;%%

%\cite{Gauntlett:2009bh}
\bibitem{Gauntlett:2009bh}
  J.~P.~Gauntlett, J.~Sonner and T.~Wiseman,
  ``Quantum Criticality and Holographic Superconductors in M-theory,''
  JHEP {\bf 1002}, 060 (2010)
  [arXiv:0912.0512 [hep-th]].
  %%CITATION = JHEPA,1002,060;%%


%\cite{Konoplya:2009hv}
\bibitem{Konoplya:2009hv}
  R.~A.~Konoplya and A.~Zhidenko,
  ``Holographic conductivity of zero temperature superconductors,''
  Phys.\ Lett.\  B {\bf 686}, 199 (2010)
  [arXiv:0909.2138 [hep-th]].
  %%CITATION = PHLTA,B686,199;%%




%\cite{Basu:2009vv}
\bibitem{Basu:2009vv}
  P.~Basu, J.~He, A.~Mukherjee and H.~H.~Shieh,
  ``Hard-gapped Holographic Superconductors,''
  Phys.\ Lett.\  B {\bf 689}, 45 (2010)
  [arXiv:0911.4999 [hep-th]].
  %%CITATION = PHLTA,B689,45;%%




\end{thebibliography}
\end{document}